\relax
\tolerance=500
\input iopppt
\jnlstyle
\title{Ideal gas sources for the Lem\^aitre-Tolman-Bondi metrics}
\author{ Roberto A. Sussman}
\address{ Instituto de Ciencias Nucleares, UNAM, Apartado Postal 70-543, 
M\'exico D.F. 04510,  M\'EXICO} 

\abs
New exact solutions emerge by replacing the dust source of
the Lem\^aitre-Tolman-Bondi metrics with a viscous fluid satisfying the monatomic
gas equation of state. The solutions have a consistent thermodynamical
interpretation. The most general transport equation of Extended
Irreversible Thermodynamics is satisfied, with phenomenological coefficients bearing a close
resemblance to those characterizing a non relativistic Maxwell-Bolzmann gas.   
\endabs

\section{Introduction.}
Dust is the prefered matter source in general relativistic models if rest mass is the
dominant form of matter-energy$^{[1,2]}$. Exact solutions with a dust source might
provide an appropriate description of the dynamics of astrophysical systems, but lack
any information on their thermal properties and evolution. A more complete description
can be achieved by the replacement of the dust source with a non-relativistic monatomic
ideal gas. Since natural processes are irreversible and ideal gases (in
general) are characterized by non-zero transport coefficients, a hydrodynamic
model of this type of source should be based on the momentum energy tensor of an
imperfect fluid.  The purpose of this paper is to present a new class of exact solutions
in which such matter tensor, subjected to the ideal gas equation of state, is the
source of the Lem\^aitre-Tolman-Bondi (LTB) metrics$^{[1,2]}$. However, Einstein field
equations applied to these metrics (in comoving coordinates) necessarily
restrict the fluid tensor to zero heat conduction and a non-accelerating 4-velocity,
so that shear viscosity, associated with an anisotropic pressure tensor, becomes the
only dissipative flux and the resulting shear viscosity tensor must be of a very
special kind: its divergence must exactly ballance a non-zero gradient of the
equilibrium pressure. Since, in general, classical ideal gases have non-zero  heat
conduction and pressure gradients produce non-zero 4-acceleration, the lack
of these features obviously limits the physical applicability of the models, but is
justified by the mathematical simplification of the field equations. Fortunately,
though, the above mentioned limitations are compensated by the obtension of exact
solutions that are thermodynamicaly consistent: the models derived in this paper are
wholy compatible with the requirements (when shear viscosity is the only dissipative
flux) of theories dealing with irreversible phenomena, such as Extended Irreversible
Thermodynamics$^{[3-10]}$ and Kinetic Theory$^{[3,11]}$. Bearing in mind their
shortcomings, these fluid models of a classical ideal gas, evolving along adiabatic
(zero heat conduction) but irreversible (non-zero viscosity) thermodynamical
processes$^{[11,12]}$ can be considered a reasonable initial improvement on dust
sources in the description of astrophysical systems in classical conditions.  

\section{Monatomic ideal gas sources.}
A non-relativistic, non-degenerated, monatomic ideal gas is characterized by
the equation of state$^{[7,9]}$ 
$$\rho=mc^2n+{3\over{2}}nkT,\qquad p=nkT\en  $$
\noindent where $\rho,n,p,T$ are matter-energy and particle number densities,
equilibrium pressure and temperature, $m,k$ are the particles' mass and
Boltzmann's constant. As mentioned in the previous section, the hydrodynamical
description of this ideal gas will be provided by solving Einstein's field equations
for the momentum-energy tensor of a fluid with dissipative fluxes
$$ T^{ab}=\rho u^au^b+ph^{ab}+\Pi^{ab}+2c^{-1}q^{(a}u^{b)}, \en\cr
h^{ab}= c^{-2}\,u^au^b+ g^{ab},\qquad u_a\Pi^{ab}=0,\qquad \Pi^a\,_a=0,\qquad
u_aq^a=0,$$     
\noindent where the equilibrium variables $\rho,p$ satisfy (1), $u^a$ is the
4-velocity of the particle frame$^{[5]}$, $\Pi^{ab}$ is the shear viscous pressure and
$q^a$ is the heat conduction vector (bulk viscosity is negligible for a classical
ideal gas$^{[3,9]}$).  This matter tensor must satisfy energy-momentum ballance, as
well as matter conservation and  entropy production laws
$$T^{ab}\,_{;b}=0\enpt\cr
(nu^a)_{;a}=0 \enpt\cr
(nsu^a)_{;a}\geq0 \endpt$$  
\noindent where $s$ is the entropy per particle, which together with $q_a$ and
$\Pi^a_b$ must comply with suitable transport and constitutive equations of
Irreversible Thermodynamics.

\section{A class of exact solutions.}
Consider the Lem\^aitre-Tolman-Bondi (LTB)
metric ansatz associated with spherically symmetric dust solutions$^{[1,2]}$
$$ds^2=
-c^2dt^2+{{Y'^2}\over{1-k_0f^2}}dr^2+Y^2\left(d\theta^2+\sin^2\theta d\phi^2
\right)\en
$$
\noindent where $k_0=0,\pm 1$, $Y=Y(t,r)$, $f=f(r)$ and $Y'=Y_{,r}$.
Following the strategy outlined previously, we shall consider the fluid tensor (2)
(instead of dust) as the matter source of (4). If the spacial coordinates in (4)
are comoving, we have $u^a=c\delta^a_t$, and so, the 4-velocity is a geodesic
field ($u_{a;b}u^b=0 $) characterized by two kinematical invariants: the expansion
scalar $\Theta= u^a\,_{;a}$ and the shear tensor $\sigma_{ab}=u_{(a;b)}-(1/3)\Theta
h_{ab}$, given explicitly by
$$\Theta={{2\dot Y}\over{Y}}+{{\dot Y'}\over{Y'}}  \enpt$$
$$\fl\sigma^a\,_b=  {\bf {\hbox{diag}}}\left[
0,-2\sigma,\sigma,\sigma\right],\qquad
\sigma_{ab}\sigma^{ab}=6\sigma^2, \qquad \sigma= 
{1\over{3}}\left( {{\dot Y}\over{Y}}-{{\dot Y'}\over{Y'}}\right)
\endpt$$

\noindent where $\dot Y=u^aY_{,a}=cY_{,t}$. The field equations for (2) and (4) imply
vanishing heat coinduction: $q_a=q(t,r)\delta_a^r=0$, while the shear viscosity tensor
in (2) (a traceless symmetric tensor) takes the form
$$
\Pi^a\,_b=  {\bf {\hbox{diag}}}\left[ 0,-2P,P,P\right],\qquad
\Pi_{ab}\Pi^{ab}=6P^2\en$$
\noindent where $P=P(t,r)$ is an arbitrary function to be determind by the
field and transport equations. If $q_a=\dot u_a=0$ the ballance equation (3a) yields
$$\dot \rho +(\rho +p)\Theta +\sigma _{ab}\Pi ^{ab}=0\quad \Rightarrow \quad \dot
p+{\textstyle{5 \over 3}}\Theta p+4\sigma P=0\enpt$$
$$h^b_a(p_{,b}+\Pi _{bc;d}h^{cd})=0\quad  \Rightarrow   \quad -p'+2P'+6P{{Y'} \over
Y}=0\endpt$$

\noindent where $\dot \rho=u^a\rho_{,a},\,\dot p=u^ap_{,a}$, the quantities $\sigma,P$
are defined in (5) and (6), and the relation $\dot n+n\Theta=0$ from (3b), as well as
(1), were used to derive the expressions in the rhs of the implication sign.  

Einstein's field equations for (2), (4) and (6), together with the integration of the
matter conservation equation (3b), yield
$$\dot Y^2={{8\pi
G}\over{3c^2Y}}\left[M+W\left({{Y_0}\over{Y}}\right)^2\right]-k_0c^2f^2 
\en\cr
\hbox{where:}\qquad M\equiv mc^2\int{n_0Y_0^2Y'_0dr},\qquad W\equiv
{3\over{2}}\int{n_0kT_0Y_0^2Y'_0dr}\en $$
\noindent and the subindex ``$_0$'' below any function (as in $n_0, T_0, Y_0
$) will  denote henceforth evaluation of the function along a suitable
intial hypersurface $t=t_0$. With the help of (8) and (9), the thermodynamical
variables and
$\sigma$ are given by 
$$n=n_0\left({{Y_0}\over{Y}}\right)^3{{Y'_0/Y_0}\over{Y'/Y}} \en\cr 
T=T_0\left({{Y_0}\over{Y}}\right)^2\Psi \en\cr 
p=n_0kT_0 \left({{Y_0}\over{Y}}\right)^5{{Y'_0/Y_0}\over{Y'/Y}}\,\,\Psi
\en\cr
P={1\over{2}}n_0kT_0
\left({{Y_0}\over{Y}}\right)^5{{Y'_0/Y_0}\over{Y'/Y}}\,\,\Omega \en\cr 
\fl{{\dot Y}\over{Y}}{{Y'}\over{Y}}\sigma =-{{4\pi
G}\over{9}}mn\left[\Gamma+{3\over 2}{{kT_0}
\over{mc^2}}\left({{Y_0}
\over{Y}}\right)^2\Omega\right]{{Y'}\over{Y}}+
{{k_0c^2f^2}\over{3Y_0^2}}\left({{Y_0}
\over{Y}}\right)^2\left({{f'}\over{f}}- {{Y'}\over{Y}}\right)
\en$$
\noindent where:
$$\fl\Psi \equiv 1+{{2W} \over {W'}}\left( {{{Y'_0} \over
{Y_0}}-{{Y'} \over Y}} \right),\qquad \Omega \equiv 1+{W \over {W'}}\left(
{{{2Y'_0} \over {Y_0}}-{{5Y'}  \over Y}} \right),\qquad \Gamma\equiv
1-{{3M}\over{M'}}{{Y'}\over{Y}}\en$$
The solutions characterized by (1)-(15) become fully determined once (8) is 
integrated for specific initial conditions prescribed by selecting $n_0, T_0,
Y_0, f$ (one of these functions can always be eliminated by an arbitrary
rescaling of $r$). Ballance equations are identicaly satisfied, this follows by
substitution of (5), (11), (12) into (7).  Two important limiting cases emerge: (a)
Lem\^aitre-Tolman-Bondi dust solutions are the zero temperature limit
($T_0=0\Rightarrow T=p=P=0$), and (b) The shear-free particular case ($\sigma=0$, for
all fluid worldlines) is the FLRW metric characterized by $f'/f=Y'/Y$, $Y=R(t)f$,
$Y_0=R_0f$, $n'_0=T'_0=0$, whose source is a perfect fluid  ($P=0$) satisfying the
equation of state (1)$^{[13]}$. 

\section{Thermodynamical consistency.}
After integrating the field equations (pending the integration of (8)), it is
necessary to examine the thermodynamical consistency of the viscous fluid
source.  For such a fluid, Extended Ireversible Thermodynamics associates a
generalized entropy current$^{[3-10]}$ satisfying (3c) and relating the
deviation from equilibrium due to viscosity
$$(nsu^a)_{;a}\geq0\Rightarrow \dot s\geq0, \qquad s=s^{(e)}-\alpha
\pi_{ab}\pi^{ab}\en$$
\noindent  where $s^{(e)}$ follows from the integration of the equilibrium Gibbs
equation ($T\d s=\d(\rho/n)+p\d(1/n)$) and $\alpha$ is a phenomenological coefficient.
The evolution of the viscous pressure is in turn described by the transport
equation$^{[8-10]}$ 
$$\tau \dot\Pi_{cd}\,h^c_ah^d_b+\Pi_{ab}\left[
1+{1\over2}T\eta\left({{\tau}\over{T\eta}}\,u^c\right)_{;c}\right]+
2\eta\,\sigma_{ab}=0\en
$$
\noindent where $\dot\Pi_{ab}\equiv\Pi_{ab;c}u^c$, and the phenomenological
quantities $\tau,\eta$ are the relaxation time and the coefficient of shear
viscosity. Equation (17) is the most general transport equation available when shear
viscosity is the only dissipative effect. In order to deal with a mathematicaly
simpler problem, less general equations are often suggested. For example, Eckart's
equation ($\tau=0$) or the ``truncated form'' without the term involving $u^c$ and $T$
in (17). However, the former leads to acausal propagation equations and unstable
solutions$^{[4,9]}$, while numerical studies in FLRW spacetimes and other
theoretical arguments$^{[8-10]}$ point to unphysical effects in truncated
forms of viscosity transport equations. 

In order to solve equations (16) and (17) for (1)-(15), the coefficients
$\alpha,\eta,\tau$ must be derived or specified. As a
convenient reference providing a guideline on how to infere physicaly reasonable forms
of these quantities, consider a classical monatomic ideal gas characterized by a
Maxwell-Boltzmann distribution near equilibrium$^{[3]}$. Grad's fourteen moments
method provides the following forms for $\alpha,\eta,\tau$ for this ideal gas
$$\fl \tau_{_{MB}} ={1 \over {\gamma_0n}}\quad(a),\qquad \eta_{_{MB}}
=p\tau_{_{MB}}\quad(b),\qquad 
\alpha_{_{MB}} ={\tau_{_{MB}} 
\over {4\eta_{_{MB}}\, n T}}={k \over {4p^2}}\qquad(c),\qquad\en$$
\noindent where $\gamma_0$ is a collision integral (eq (3.41) of [3]). A reasonable
test of the physical viability of the solutions follows from verifying if the
forms of $n$, $T$, $p$, $\Pi_{ab}$ and $\sigma_{ab}$ in (5), (6), (11)-(15) are
compatible with equations (16) and (17) in which the phenomenological
coefficients are given by (18). This is examined below.
  
Substitution of (11)-(13) into (16), with $\alpha$ given by (18c) and
$s^{(e)}$ obtained from the integration of the equilibrium Gibbs equation for (1),
yields 
$$\fl s=s_0+k\ln\left[\left({{T}\over{T_0}}\right)^{3/2}
{{n_0}\over{n}}\right]-{3\over{2}}k\left({{P}\over{p}}\right)^2=s_0+k
\ln\left[ {{Y'/Y}\over{Y'_0/Y_0}}\,
\Psi^{3/2}\right]-{3\over{8}}k\left({{\Omega}\over{\Psi}}\right)^2
\en
$$
\noindent where $s_0(r)$ is an arbitrary initial value of $s$. Regarding
(17), if we assume: $\eta=p\,\tau$, with $p$ given by (12), 
then this transport equation is satisfied for $\tau$ given by:  
$$\fl\tau=-{{\Omega\Psi} \over
{4\sigma}}\left[\left[{2W\over W'}{Y'\over Y} -{7\over
16}\left(1+{2W\over W'}{Y'_0\over Y_0}\right)\right]^2+{207\over
256}\left(1+{2W\over W'}{Y'_0\over Y_0}\right)^2\right]^{-1}\en$$
While the form of $\eta$ can be justified as being formaly  identical to the
Maxwell-Boltzmann form (18b) (a remarkable fact), the form of $\tau$ in (20) is
acceptable as long as this expression behaves as a relaxation parameter for the system,
namely: it should be a positive function and its relation with $\dot s$
(computed from (19)) must be such that it garantees $\dot s\geq 0$ along
$u^a$ for all fluid worldlines. Idealy, of course, $\tau$ should be related to the
mean collision time of the gas and should also, somehow, relate to or approach the 
Maxwell-Boltzmann form (18a). Evaluating  $\dot s$ from (19) and comparing with (20),
we find the following simple and elegant relation between
$\dot s$ and
$\tau$ 
$$\dot s={3\over
4}\,k\,\left({\Omega\over\Psi}\right)^2\,{1\over\tau}=3
\,k\,\left({P\over p}\right)^2\,{1\over\tau}\en  
$$
which implies: $\dot s>0\Leftrightarrow \tau>0 $. This relation is in fact a
direct consequence of the general expression: $\dot s=\Pi_{ab}\Pi^{ab}/(2\eta nT)$
(see [4], eqn. (18)), together with equations (3b), (3c) and (18b). Therefore, from
equations (11) and (19)-(21), the following are necessary and sufficient conditions
for 
$T>0,\tau>0,\dot s>0$ 
$$ \Psi>0\enpt\cr \sigma\Omega<0\enpt$$
\noindent while the conditions insuring that $\dot s$ decreases for increasing $\tau$ 
($\dot\tau>0 \Leftrightarrow \ddot s<0$) follow from (20)-(21)  
$$\dot \tau>0,\qquad {\ddot s\over\dot s}={{9\sigma}\over{\Psi\Omega}}\,{W\over
W'}{Y'\over Y}\left(1+{2W\over W'}{Y'_0\over
Y_0}\right)-{\dot\tau\over{2\tau}}<0\endpt$$

\noindent  The set (22) provides the necessary and sufficient conditions for a
theoreticaly consistent thermodynamical description of the solutions, within the
framework of equations (16)-(21). However, in order to work out these conditions in a
more explicit and direct manner (especialy (22c)), we need to know the explicit form of
$Y'/Y$ from integrating (8).       

\section {The case $k_0=0$.}
Consider the integration of (8). For the simplest case, $k_0=0$ (the cases $k_0=\pm 1$
are examined in [14]), this yields
$$\fl {3 \over 2}\sqrt {{{8\pi GM} \over {3c^2Y_0^3}}}\left(  {t-t_0}
\right)=\sqrt {y\left( {y^2+\delta } \right)}-\sqrt {1+\delta
}+{{\delta ^{3/ 4}} \over 2}\left( {\hbox{F}}-{\hbox{F}_0}
\right)\enpt\cr
\fl {{Y'} \over Y}=\cr
\fl {{M'} \over {3M}}\left\{ {1-\tilde\Gamma {{\sqrt 
{y^2+\delta }} \over {y^{5/ 2}\sqrt {1+\delta }}}+{9 \over
4}{{kT_0} \over {mc^2}}\tilde\Omega \left[ {{{\sqrt {y^2+\delta }} \over
{y^{5/ 2}}}\left( {{{2\delta } \over {\sqrt {1+\delta
}}}-\delta ^{3/ 4}\left( {\hbox{F}}-{\hbox{F}_0}\right)} \right)-{{2\delta }
\over {y^2}}} \right]} \right\}\endpt
\cr
\fl\hbox{where:}\qquad \tilde\Gamma\equiv 1-{3M\over M'}{Y'_0\over Y_0},\qquad
\tilde\Omega\equiv 1+{W\over W'}\left({2Y'_0\over Y_0}-{5\over 3}{M'\over M}
\right)
$$
\noindent $\delta\equiv W/M$, $y\equiv Y/Y_0$ and ${\hbox{F}}$ is the
elliptic integral of the first kind with modulus $1/\sqrt{2} $ and argument
$\varphi$ given by: $\cos\varphi=(y-\sqrt{\delta})/(y+\sqrt{\delta}) $. With the help
of (23b) and the definitions of $\Psi,\Omega,\sigma$, it is possible to provide
specific examples showing that conditions (22) are satisfied for physicaly reasonable
initial value functions  $n_0, T_0, Y_0$. Consider the following trial functions,
analogous to empirical models of mass distribution in spherical gallaxies and
clusters$^{[15,16]}$   
$$\fl n_0=\bar n_0\left[1+{{a}\over{1+(Y_0/R_0)^2}}\right],\qquad T_0=\bar
T_0\left[1+{{b}\over{1+(Y_0/R_0)^2}}\right], \qquad Y_0=R_0r\en$$ 
\noindent where $\bar n_0,\bar T_0,a,b$ are positive empirical constants,  while $R_0$
is a characteristic length. Let us assume that (24) provides the initial particle
number density and temperature profile at the time when matter and radiation decouple,
$t_0=t_{dec.}$, of a spherical inhomogeneity of gallactic mass ($10^{44}$ gm) in an
expanding FLRW background characterized by an intial profile:
$\bar n_0\approx10^5\hbox{cm}^{-3}$ and $\bar T_0\approx 3\times 10^3\hbox{K}$. 
Taking $m$ to be a protonic mass and considering a red shift factor $z\approx 10^3$
(so that $R_0\approx 10^{20}$cm), reasonable estimates of $a,b$ in (24) are $a\approx
10^3$ and
$b\approx 10$. As illustrated by figures (1a), (1b) and (1c) below, we have positive and
monotonously decreasing $T$ and $\dot s$, and monotonously increasing $\tau$, so that
the evolution of a configuration of this type is consistent with the thermodynamical
requirements (16)-(22). In fact, it is possible to show that conditions (22) hold for
all trial functions of the form (24), provided
$a,b$ comply with the constraint
$${2\over 3}\left(1+{1\over b} \right)<1+ {1\over a}\en
$$
\noindent together with $kT_0/mc^2<1$, $Y\geq Y_0$, conditions satisfied
by a large choice of empirical constants. For $n_0,T_0$ given by  (24) and (25), we
also have$^{[14]}$: $\tau\propto 1/n$ for $Y/Y_0\gg 1$, a form of $\tau$ qualitatively
analogous to (18a) and to relaxation times of globular clusters$^{[17]}$. Further
extension and development of the work presented in this paper is being
undertaken.$^{[14]}$.

\ack
I acknowledge finantial support from CONACYT grant 3567E. I am grateful to H. Quevedo, J.
Triginer and D. Pav\'on for useful suggestions and encouragement. Acknowledgement  is
also due to the memory of Deedee and
Chocho, whose wise meowing contributed to clarify the author's ideas.

\bigskip
\noindent{\bf Figure Caption.}
The figures display: $\dot s$ (Figure (1a)), $T$ (Figure (1b)) and $\log_{10}(\tau)$
(Figure (1c)), ploted vs.
$\arctan(r)$ and $\arctan(y)$ (for $y\geq1$, where $y\equiv Y/Y_0$), for the spherical
inhomogeneity described in the text. Notice that $\dot s$ and $\tau$ are everywhere
positive and they respectively decrease/increase along increasing $y$ (roughly
equivalent to the fluid worldlines). This example is not meant to be a model of a
``real'' gallaxy, but simply to illustrate that conditions (22) hold for physicaly
reasonable initial value functions
$n_0,T_0$. These plots were done with the symbolic package MapleV. 

\numreferences 

\rf{[1]} Kramer D, Stephani H, MacCallum MAH., Herlt E 1980. {\it{Exact
solutions of Einstein's field equations}}. Cambridge University Press,
Cambridge, U.K. See also: Krasi\'nski A 1997. {\it{Inhomogeneous
cosmological models}}.  Cambridge University Press, Cambridge, U.K.
\rf{[2]}  Dyer CC 1976, {\it{MNRAS}},
{\bf 175}, 420. Raine DJ and Thomas EG 1981, {\it{MNRAS}}, {\bf 195},
649.  Panek M 1992, {\it{Ap. J.}}, {\bf 388}, 225. Saez D, Arnau JV and
Fullana MJ 1983, {\it{MNRAS}}, {\bf 263}, 681.  
\rf{[3]} Jou D, Casas V\'azqez J and Lebon G 1993. {\it{Extended
Irreversible Thermodynamics.}}.  Springer Verlag, Berlin, Heidelberg, New
York.
\rf{[4]} Hiscock W and Lindblom L 1983, {\it{Ann. Phys. (NY).}}, {\bf 151},
466. 
\rf{[5]} Israel W 1976, {\it{Ann. Phys. (NY).}}, {\bf 100}, 30.
See also: Israel W 1989, in {\it{Relativistic Fluid Dynamics.}} Eds. M. Anile
and Y. Choquet-Bruhat, Spriger-Berlin.
\rf{[6]} Jou D, Casas V\'azqez J, Lebon G 1988, {\it{Rep. Prog. Phys.}},
{\bf 51}, 1104. See also: Pav\'on D, Jou D and Casas V\'azqez J 1982,
{\it{Ann. Inst. Henri Poincare A}}, {\bf 36}, 79.
\rf{[7]} Muller I and Ruggieri T 1993, {\it{Extended Thermodynamics.}},
Springer Tracts in Natural Philosophy, Volume 37. Springer Verlag, Berlin,
Heidelberg, New York.
\rf{[8]} R. Maartens 1995, {\it{Class. Quantum Grav.}},
{\bf 12}, 1455.
\rf{[9]} Hiscock WA and Salmonson J 1991, {\it{Phys. Rev. D}},
{\bf 43}, 3249.   
\rf{[10]} Pav\'on D and Romano V 1993,  
{\it{Phys. Rev. D}}, {\bf 47}, 1396. See also: M\'endez V and
Triginer J 1996, {\it{J. Math. Phys.}}, {\bf {37}}, 2906.
\ref{[11]} G. Wannier 1987. {\it{Statistical Physics.}}.  Dover
Publications Inc., New York.
\rf{[12]} Zemansky MW 1957. {\it{ Heat and Thermodynamics.}}. MaGraw-Hill
Book Company, New York, Toronto, London. See chapter 8.
\rf{[13]} Calv\~ao MO and Lima JAS 1989, {\it{Phys. Lett. A}},
{\bf 141}, 229. See also: Garecki J and Stelmach J 1990, 
{\it{Annals of Physics}}, {\bf 204}, 315.
\rf{[14]} Sussman RA and Triginer J 1998. In preparation. 
\rf{[15]} I. King 1962, {\it{The Astronomical Journal}},
{\bf 67}.
\rf{[16]} {\it{Galaxies, Quasars and Cosmology.}}. L.Z. Fang and R.
Ruffini, Editors 1985. Advanced Series in Astrophysics and Cosmology, vol. 2.
World Scientific Publishing Company, Pte. Ltd. See pp.
50-52.  
\rf{[17]} {\it{Relativity Theory and Astrophysics. 2. Galactic
Structure.}}. J. Ehlers, Editor 1967. Lectures in Applied Mathematics, vol 9.
American Mathematical Society. Providence, Rhode Island. See pp.
44-53.
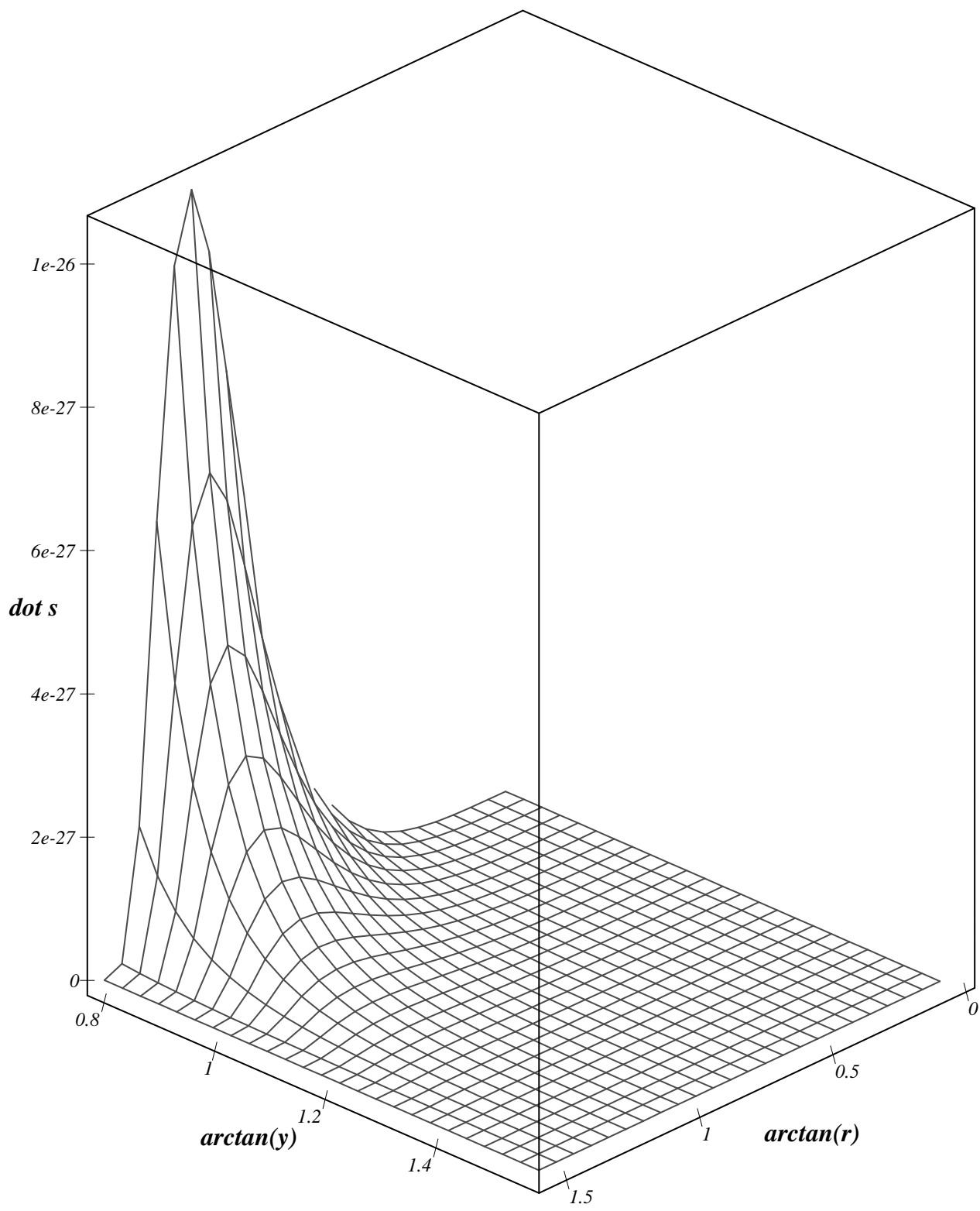
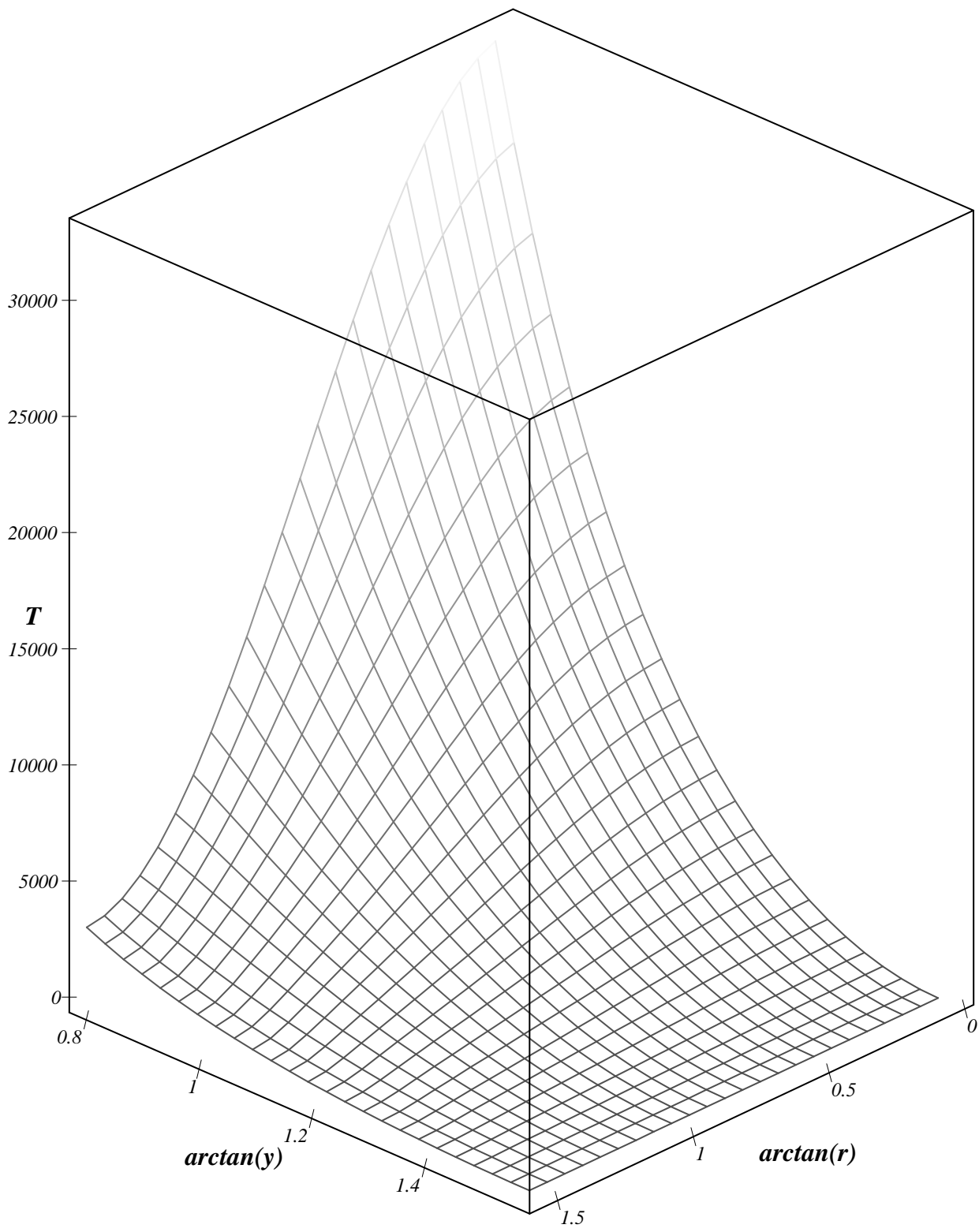
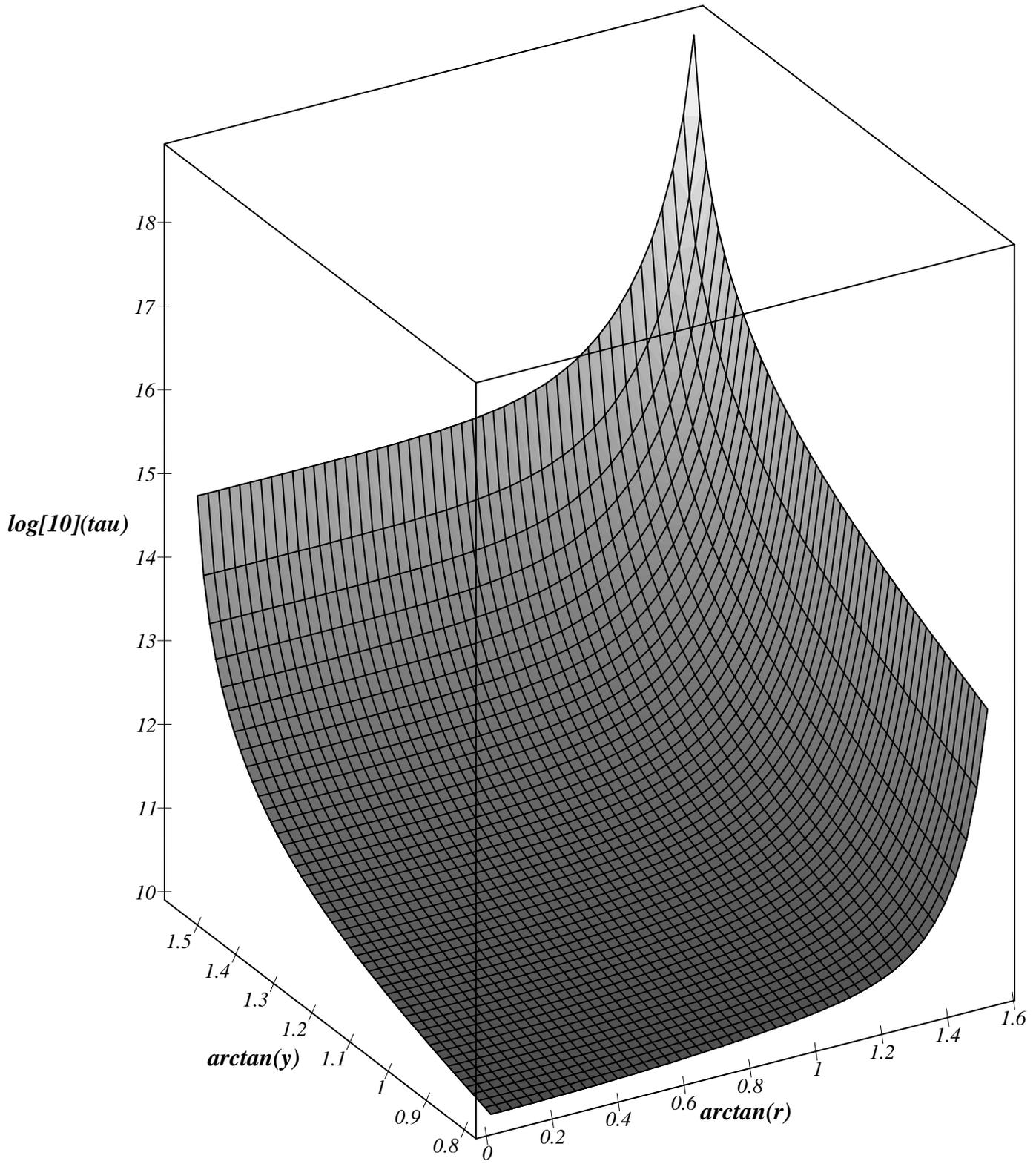
\bye